\documentclass[conference]{IEEEtran}
\IEEEoverridecommandlockouts
\usepackage{cite}
\usepackage{amsmath,amssymb,amsfonts}
\usepackage{algorithmic}
\usepackage{titlesec}
\usepackage{graphicx}
\usepackage{textcomp}
\usepackage{xcolor}
\usepackage{fancyhdr}
\usepackage{url}
\usepackage{multirow}
\usepackage{booktabs}
\usepackage{gensymb} 
\usepackage{caption}
\usepackage[font=footnotesize,skip=5pt]{caption}
\usepackage{subcaption}
\usepackage{float} 
\usepackage{array} 
\usepackage{balance}

\usepackage[nolist,nohyperlinks]{acronym}
\def\BibTeX{{\rm B\kern-.05em{\sc i\kern-.025em b}\kern-.08em
    T\kern-.1667em\lower.7ex\hbox{E}\kern-.125emX}}
\begin{acronym} 
\acro{3GPP}{3rd Generation Partnership Project}
\acro{ACF}{auto correlation function}
\acro{ADC}{analog-to-digital converter}
\acro{AWG}{arbitrary waveform generator}
\acro{BEM}{basis-expansion model}
\acro{BPF}{band-pass filter}
\acro{CE}{complex exponential}
\acro{CDF}{cumulative density function}
\acro{CIR}{channel impulse response}
\acro{CP}[CP]{cyclic prefix}
\acro{CTF}{channel transfer function}
\acro{D2D}{device-to-device}
\acro{DAC}{digital-to-analog converter}
\acro{DC}{direct current}
\acro{DFT}{discrete Fourier transform}
\acro{DOD}{direction of departure}
\acro{DOA}{direction of arrival}
\acro{DPS}{discrete prolate spheroidal}
\acro{DPSWF}{discrete prolate spheroidal wave function}
\acro{DSD}{Doppler spectral density}
\acro{DSP}{digital signal processor}
\acro{DR}{dynamic range}
\acro{ETSI}{European Telecommunications Standards Institute}
\acro{FIFO}{first-input first-output}
\acro{FFT}{fast Fourier transform}
\acro{FMCW}{frequency modulated continuous wave}
\acro{FP}{fixed-point}
\acro{FPGA}{field programmable gate array}
\acro{GSCM}{geometry-based stochastic channel model}
\acro{GSM}{global system for mobile communications}
\acro{GPS}{global positioning system}
\acro{HPBW}{half power beam width}
\acro{ICI}[ICI]{inter-carrier interference}
\acro{IDFT}{inverse discrete Fourier transform}
\acro{IF}{intermediate frequency}
\acro{IFFT}{inverse fast Fourier transform}
\acro{IQ}{in-phase and quadrature-phase}
\acro{ISI}{inter-symbol interference}
\acro{ITS}{intelligent transportation system}
\acro{MEC}{mobile edge computing}
\acro{MSE}{mean square error}
\acro{LLR}{log-likelihood ratio}
\acro{LO}{local oscillator}
\acro{LOS}{line-of-sight}
\acro{LMMSE}{linear minimum mean squared error}
\acro{LNA}{low noise amplifier}
\acro{LPF}{low-pass filter}
\acro{LSF}{local scattering function}
\acro{LTE}{long term evolution}
\acro{LUT}{look-up table}
\acro{LTV}{linear time-variant }
\acro{MIMO}{multiple-input multiple-output}
\acro{MPC}{multi-path component}
\acro{MC}{Monte Carlo}
\acro{NI}{National Instruments}
\acro{NLOS}{non-line of sight}
\acro{OFDM}{orthogonal frequency division multiple access}
\acro{OTA}{over-the-air}
\acro{PA}{power amplifier}
\acro{PAPR}{peak-to-average power ratio}
\acro{PC}{personal computer}
\acro{PDP}{power delay profile}
\acro{PER}{packet error rate}
\acro{PLL}{phase-locked loop}
\acro{PPS}{pulse per second}
\acro{QAM}{quadrature ampltiude modulation}
\acro{QPSK}{quadrature phase shift keying}

\acro{RB}{resource block}
\acro{RBP}{resource block pair}
\acro{RF}{radio frequency}
\acro{RMS}{root mean square}
\acro{RSSI}{receive signal strength indicator}
\acro{RT}{ray tracing}
\acro{RX}{receiver}
\acro{SCME}{spatial channel model extended}
\acro{SDR}{software defined radio}
\acro{SISO}{single-input single-output}
\acro{ReRoMA}{Redirecting Rotating Mirror Arrangement}
\acro{ISAC}{integrated sensing and communication}
\acro{BW}{bandwidth}
\acro{RSU}{roadside unit}
\acro{STFT}{short-time Fourier transform}
\acro{SIW}{substrate integrated waveguide}
\acro{SoCE}{sum of complex exponentials}
\acro{SNR}{signal-to-noise ratio}
\acro{SUT}{system-under test}
\acro{SSD}{soft sphere decoder}
\acro{TBWP}{time-bandwidth product}
\acro{TDL}{tap delay line}
\acro{TX}{transmitter}
\acro{UMTS}{universal mobile telecommunications systems}
\acro{UDP}{user datagram protocol}
\acro{URLLC}{ultra-reliable and low latency communication} 
\acro{US}{uncorrelated-scattering}
\acro{USRP}{universal software radio peripheral}
\acro{VNA}{vector network analyzer}
\acro{ViL}{vehicle-in-the-loop}
\acro{V2I}{vehicle-to-infrastructure}
\acro{V2V}{vehicle-to-vehicle}
\acro{V2X}{vehicle-to-everything}
\acro{VST}{vector signal transceiver}
\acro{VTD}{virtual test drive}
\acro{WF}{Wiener filter}
\acro{WSS}{wide-sense-stationary}
\acro{WSSUS}{wide-sense-stationary uncorrelated-scattering}
\acro{WAVE}{wireless access in vehicular environments}
\acro{USC}{University of Southern California}
\acro{BUT}{Brno University of Technology}
\acro{AIT}{Austrian Institute of Technology}
\acro{TUW}{Technical University Vienna}
\acro{cmWave}{centimeter wave}
\acro{mmWave}{millimeter wave}
\acro{WLAN}{wireless local area network}
\acro{NR}{new radio}
\acro{PN}{pseudo-noise}
\acro{Rb}{rubidium}
\acrodef{SoC}{system on chip}
\acrodef{RHCP}{right hand circular polarization}
\acrodef{LHCP}{left hand circular polarization}
\acrodef{RAID}{redundant array of independent disks}
\acro{VW}{Volkswagen}
\acro{HRPE}{high resolution parameter estimation}
\acro{MMSE}{minimum mean square error}
\end{acronym}

\fancypagestyle{firstpage}
{
    \fancyhf{}
    \fancyhead[C]{The paper has been presented at the 2025 IEEE Future Networks World Forum (FNWF),\\ Bangalore, India, November 10-12, 2025, \url{https://doi.org/10.1109/FNWF66845.2025.11317288}}
    \fancyfoot[C]{This research was funded in part by the National Science Center (NCN), Poland, grant no. 2021/43/I/ST7/03294 (MubaMilWave). For this purpose of Open Access, the author has applied a CC-BY public copyright license to any Author Accepted Manuscript (AAM) version arising from this submission.}
}

\begin{document}

\title{Comparison of 60 GHz and 80 GHz Vehicle-to-Vehicle Channels Using Delay and Doppler Characteristics\\

} 

\author{
    \IEEEauthorblockN{
    Ales Prokes\IEEEauthorrefmark{1}, 
    Tomas Mikulasek\IEEEauthorrefmark{1},
    Josef Vychodil\IEEEauthorrefmark{1},  
    Radek  Zavorka\IEEEauthorrefmark{1},
    Jiri Blumenstein\IEEEauthorrefmark{1}, \\ 
    Jaroslaw Wojtun\IEEEauthorrefmark{3}, 
    Jan M. Kelner\IEEEauthorrefmark{3}, 
    Cezary Ziolkowski\IEEEauthorrefmark{3}, 
    Aniruddha Chandra\IEEEauthorrefmark{2}}
    
    \IEEEauthorblockA{
        \IEEEauthorrefmark{3}Institute of Communications Systems, Military University of Technology, Warsaw, \textit{Poland};\\
        \IEEEauthorrefmark{2}ECE Department, National Institute of Technology, Durgapur, \textit{India};\\
        \IEEEauthorrefmark{1}Brno University of Technology, Brno, \textit{Czechia}, Email: prokes@vut.cz
    }
}
\titlespacing*{\subsection}
{0pt}{*0.7}{*0.3}
\maketitle

\titlespacing*{\section}
{0pt}{*0.7}{*0.3}
\maketitle

\setlength{\belowcaptionskip}{-10pt}

\maketitle
\thispagestyle{firstpage}

\begin{abstract}
The aim of this paper is to provide a comparison of channel characteristics for vehicle-to-vehicle (V2V) communication at 60 GHz and 80 GHz frequency bands in a high-mobility scenario where two vehicles pass each other in opposite directions. The study is based on measurements of the time-varying channel impulse response capturing the behavior of multi-path propagation during vehicle motion. By directly comparing these two frequency bands under identical measurement conditions, we attempt to quantify the differences in power delay profile, \ac{RMS} delay spread, \ac{RMS} Doppler spread, and intervals (regions) of stationarity in time domain. The results show that these bands do not differ significantly, but the 80 GHz band exhibits somewhat greater \ac{RMS} delay spread and \ac{RMS} Doppler spread when calculated over the entire delay-Doppler spectrum, and conversely exhibits shorter stationarity regions. However, the characteristics of the measurement setup in the two bands and their influence on comparative measurements must be considered. In particular, attention must be paid to the impact of antennas.

\end{abstract}



\section{Introduction}
\setlength{\parindent}{1em}
\setlength{\parskip}{0.5em}
The demand for higher data rates, lower latency, and greater reliability in wireless communications has driven the exploration of new frequency bands and novel communication paradigms. \Ac{mmWave} bands, typically defined between 30~GHz and 300~GHz, offer large unused bandwidths and have been identified as a key enabler for next-generation wireless systems, including 5G and beyond. However, signal propagation in mmWave bands poses significant challenges due to higher free-space path loss, sensitivity to blockages, limited diffraction, and fast temporal variations in dynamic environments. Accurate modeling of the radio channel is therefore essential to understand and mitigate these effects, especially in applications where reliability and performance are tightly coupled to the physical environment.

A particularly challenging application is high-mobility \ac{V2V} communication, where both the transmitter and receiver located on vehicles are moving at high relative speeds, causing rapid channel changes, Doppler effects, and time-varying multi-path propagation. These effects are especially pronounced at \ac{mmWave} frequencies, where channel behavior strongly depends on vehicle geometry, trajectory, antenna orientation, and surrounding objects.

Several recent works have focused on characterizing \ac{V2V} channels in the \ac{mmWave} bands based on real measurements. For example, \cite{Molisch_millimeter} provides a detailed overview of the V2X channel analyzes in various scenarios (passing vehicles, overtaking, convoy driving, and driving in urban environments), describing sounder setups and comparing key channel parameters such as delay, Doppler, and angular spread, path loss, stationarity regions, and others. It highlights how channel characteristics relate to different driving situations. 
In \cite{Wang_fading}, V2V channel measurements at 73~GHz were performed in an urban environment with vehicles approaching at 60~km/h. The study analyzes large-scale and small-scale fading, showing that the fast-fading distribution evolves from Rician to Nakagami and eventually to lognormal as the Tx–Rx distance increases.   
Similarly, the authors in \cite{hammoud2024TVT} provides real traffic measurements using a double-directional 60~GHz sounder  ReRoMA in scenarios such as convoys, oncoming traffic and overtaking and presents key channel parameters such as the path loss exponent (1.9), the \ac{RMS} delay spread (5–110~ns), and angular spread (0.05–0.4~rad). 
Extensive 41~GHz V2V measurements in urban and underground parking scenarios examined vehicle obstructions are presented in \cite{Liu_Measurements}. A stochastic path loss model is proposed for both \ac{LOS} and obstructed \ac{LOS}. The influence of certain obstacles is being investigated. For example, an obstacle in the form of truck adds 9–19~dB path loss and reduces shadow fading deviation by 1–2~dB, depending on distance to the RX. A new V2V path loss model based on 59.6~GHz measurements in crossing-vehicle scenarios, offering improved accuracy over traditional models for both approaching and receding vehicles is published in \cite{Ghosh_V2V}. 
Paper \cite{Chopra_A_Real_Time} presents a 1~GHz-wide V2V channel sounder operating at 28~GHz, using phased arrays to capture 116 directional beams in less than 1~ms.  When measured on highways and in urban environments, it revealed rich scattering, directional diversity, and waveguide-like effects that increase the reliability and range of \ac{mmWave} V2V. Ref. \cite{Hoellinger_V2V} presents initial V2V measurement results at 26~GHz in urban and suburban areas with different movement scenarios. The analysis includes time-varying power delay profiles, path loss, and delay spread, which ranges from 25 to 150~ns depending on the environment. Another study \cite{Takahashi_Distance} measures 60~GHz path loss between vehicles on urban highways and roads, showing distance-dependent attenuation that follows a 2nd‑order power law on highways and 1st‑order on city streets. It reports extra losses of ~15~dB on highways and ~5~dB on local roads beyond 30~m and confirms \ac{V2V} communication feasibility out to beyond 100~m.
Multiband V2V measurements across several frequency bands (6.75 GHz, 30 GHz, and 60 GHz) in a street canyon environment is presented in \cite{Dupleich_Multi-Band}. It analyses path loss, delay spread, and angular spread in different traffic conditions. It highlights how channel metrics vary with frequency, revealing richer scattering and reduced coherence at higher bands. In \cite{Dupleich_Multi-Band_T} the authors provide \ac{V2V} propagation data measured at a T‑intersection across the same frequencies as above, showing strong frequency-dependent behavior in path loss and multi-path richness due to building and vehicle blockage. In paper \cite{groll2019sparsity}, a 60 GHz \ac{V2I} radio channel in a suburban environment is characterized by \ac{CTF}. The transmitter was mounted on a moving vehicle. Doppler spread function demonstrated the effects of nearby parked vehicles. Propagation loss between two moving vehicle at 60~GHz was studied in \cite{yamamoto2008path} with up to three commercial vehicles blocking the LOS. Models were developed to represent NLOS propagation characteristics.
V2V measurements across 6.75, 30, 60, and 73~GHz in urban and highway settings focusing on vehicle blockage showed that blocking vehicle causes 5.5–17~dB loss, and increased delay and angular spreads depending on blocker size \cite{boban2019multi}. Note that there are other similar works carried out by teams such as NYU Wireless.

Although these studies provide valuable insights, they often focus on  static or less mobile configurations, or concentrate on a single frequency band, which limits the generality of their conclusions. A closer study of the above-mentioned and other works would reveal that the channel sounders used for measurements often have a limited bandwidth not exceeding 1~GHz, which limits the time resolution of the measurements, and have a small memory depth for recordings lasting barely a few seconds. Channel analyses are dominated by path loss characteristics influenced by distance or the environment (blocking objects). Small-scale channel features such as \ac{RMS} delay spread, angle spread, and statistical properties such as \ac{PDP} or \ac{CIR} fitting with different distributions are sometimes presented, but characteristics such as \ac{RMS} Doppler spread or stationarity regions are rarely mentioned, even though they are very important for understanding channel variability over time, which is critical for system design, channel estimation, and beamforming efficiency.

This paper aims to contribute to this area by providing  
\vspace{-5pt}
\begin{itemize}
\item \textit{Comparative analysis at 60 GHz and 80 GHz \ac{mmWave} bands}:
The study examines V2V channel behavior under identical measurement conditions in a high-mobility scenario with vehicles passing in opposite directions.

\item \textit{Long-time \ac{CIR} analysis with excellent delay-domain resolution}: Channel measurements with 2 GHz bandwidth provide detailed analysis of multi-path propagation, and large memory depth enables examination of channel characteristics evolution in delay and Doppler domains.          

\item \textit{Quantitative comparison of key parameters}:
The paper compares \ac{RMS} delay spread, \ac{RMS} Doppler spread, and stationarity regions, under identical conditions to highlight frequency-specific differences.
\end{itemize}

The rest of the paper is structured as follows: Section 2 provides an overview of the measurement scenarios. Section~3 briefly describes the measurement setup. Section 4 outlines the methods used for signal processing. Section 5 presents the results of the measurements, simulations and the calculation of channel parameters. Finally, the conclusion summarizes the key findings of the paper.

\section{Measurement scenario}

\begin{figure}
    \centering
    \includegraphics[width=0.975\linewidth]{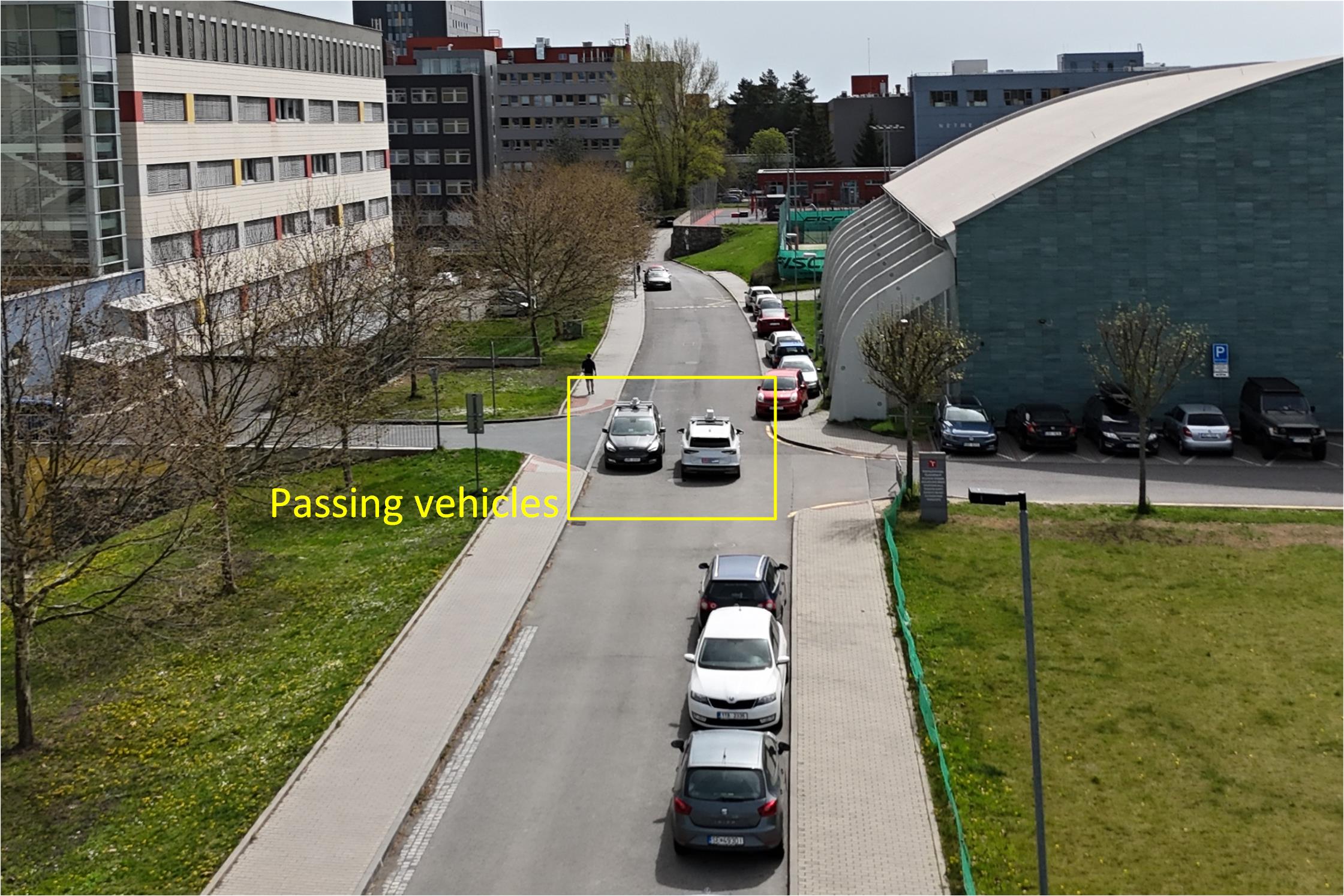}
    \caption{Measurement \ac{V2V} scenario for passing vehicles in the opposite direction at BUT}
    \label{fig:Scenario_1}
    \vskip 3pt
\end{figure}

\begin{figure}
    \vspace{-0pt}
    \centering
    \includegraphics[width=0.975\linewidth]{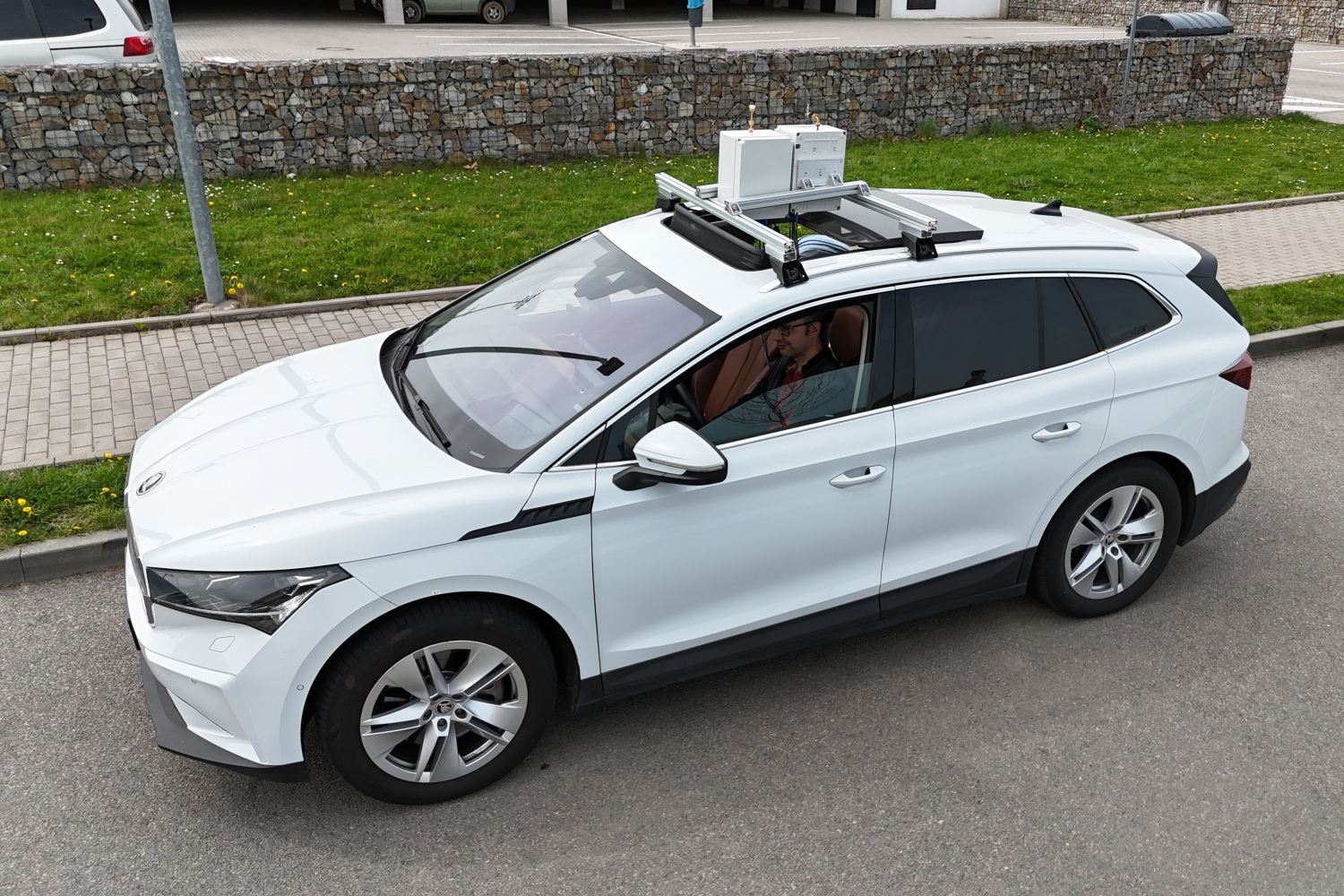}
    \caption{Measuring vehicle with channel sounder transmitters mounted on the roof of the vehicle}
    \label{fig:Car_mount_1}
    \vskip-5pt
\end{figure}

The measurement of passing vehicles was carried out on the campus of the Faculty of Electrical Engineering and Communication Technologies of the Brno University of Technology between the building at Technická 12, the sports complex, and the parking lot (see Fig.~\ref{fig:Scenario_1} and https://mapy.com/s/goculagone). Near the road within the measurement range, there are parked cars, streetlights, traffic signs, medium-sized trees, and sports complex building.  The pillars and wall of the building are approximately 3 m and 6.5 m away from the road. The wall of the building is 9 m high. The road is approximately 6 m wide and the sidewalks on both sides are 2 m wide. Omnidirectional antennas and outdoor units for channel measurement were mounted on the roofs of the vehicles. The location of the transmitting units on the Škoda Eniaq car is shown in Fig.~\ref{fig:Car_mount_1} (the 80 GHz unit is located at the front). The receiving units of the sounder are mounted similarly to the Ford Galaxy car.

In this paper, we present the results of two measurements of passing cars. During the first measurement, the cars passed each other at the location shown in the Fig.~\ref{fig:Scenario_1}. During the second measurement, the passing location was approximately 20 meters further away, i.e., closer to the sports complex. In the first measurement, the speed of the cars was approximately 30 km/h and 50 km/h, and in the second measurement it was approximately 35 km/h and 45 km/h. The first measurement is described in detail, including graphs and an overview of selected parameters.  Due to limited space, the second measurement is presented only as the overview of channel parameters.

\begin{table*}[ht]
\centering
\caption{Comparison of channel sounder component parameters for the 60 GHz and 80 GHz bands}
\resizebox{\textwidth}{!}{%
\begin{tabular}{>{\raggedright\arraybackslash}p{3.75cm} c c c c c c}
\toprule
& \multicolumn{3}{c}{\textbf{60 GHz}} & \multicolumn{3}{c}{\textbf{80 GHz}} \\
\cmidrule(lr){2-4} \cmidrule(lr){5-7}
 & PA & LNA & Antenna & PA & LNA & Antenna \\
\midrule[\heavyrulewidth]
Model & QPW-50662330-C1 & QLW-50754530-I2 & Cone-monopole (*) & Cerus 4 AA015 & LNF-LNR55\_96WA\_SV & SAO-6039030230-12-S1 \\
Frequency range [GHz] & 50--66 & 50--75 & 50--67 & 81--86 & 55--96 & 60--90 \\
Gain [dB, dBi] & 30 & 31 & 0.6 & 21 & 27 & 2.0 \\
Gain flatness/variation & ±3 & ±3.5 & ±1.5 & -- & ±2.5 (**) & ±3.5 \\
P1dB [dBm] (***)& 23 & -- & -- & 29 & -- & -- \\
Transmitted power [dBm] & 16 & -- & -- & 22 & -- & -- \\
Noise figure [dB] & -- & 4.5 & -- & -- & 2.86 & -- \\
3 dB Vertical Beam-width & -- & -- & 130\degree & -- & -- & 30\degree \\
\bottomrule
\vspace{1mm}
\end{tabular}%
}
\vspace{1mm}
\parbox{\textwidth}{\raggedright\footnotesize (*) Own design and manufacture. (**) For frequency above 56 GHz. (***) 1 dB compression point.}
\vspace{-5mm}
\label{tab:components}
\end{table*}

\vspace{-5pt}
\section{Measurement setup}
The 60 GHz and 80 GHz bands were used due to the availability of components (antennas, amplifiers, and converters) and because they are the subject of research in projects 23-04304 and MubaMilWave (see Acknowledgement).

The channel sounder  employs a Xilinx Zynq UltraScale+ RFSoC ZCU111 board as the base-band transmit subsystem for the both bands. It generates a \ac{FMCW} signal with ramp-up and ramp-down in order to make the spectrum as flat as possible. The  8~s  duration of \ac{FMCW} segment enables up to 125.000 measurements per seconds and the bandwidth 2.048 GHz offers distance resolution of 15 cm. Using SiversIMA FC1003E converters, this signal is up- and down-converted between baseband and the 60 GHz and 80 GHz bands. The transmitted signals are amplified by Filtronic Cerus 4 \acp{PA} at 80~GHz band and Quinstsar (product line  QPW) at 60~GHz band. The 60~GHz and 80~GHz received signals are amplified by \acp{LNA} Quinstar (product line QLW) and Low Noise Factory (product line LNF-LNR) respectively. The TD testbed has been optimized to significantly improve the sensitivity, IQ imbalances and dynamic range  \cite{vychodil2019iet}, which is about 45 dB. The \acp{TX} and \acp{RX} are synchronized by Stanford Research Systems FS740 GPS/Rubidium reference clocks. For more information on the channel sounder architecture for 80 GHz and about the preprocessing of the \ac{FMCW} signal response to obtain the CIR, see \cite{Zavorka_Charact}. Note that the 60 GHz channel sounder has the same architecture and uses similar components. 

In order to justify any differences in the analysis results for both bands, it is necessary to consider the effects of key components in both frequency sections of the channel sounder. An overview of these components can be found in Table~\ref{tab:components}. 

\section{Data processing methodology}

The basic function on which all further analyses are based and which is provided by the channel sounder is time-varying channel impulse response $h(t, \tau)$. It can be modeled as a sum of $L$ multi-path components, where each path is characterized by a time-dependent amplitude $a_l(t)$, a phase shift $\phi_l(t)$, and a delay $\tau_l(t)$.
\begin{equation}
h(t, \tau) = \sum_{l=1}^{L} a_l(t) \, e^{j\phi_l(t)} \, \delta[\tau - \tau_l(t)].
\end{equation}
\noindent
The function $\delta(.)$ represents the Dirac delta. From the \ac{CIR} we can simply get the \textit{instantaneous} \ac{PDP}
\begin{equation}
P(t, \tau) = \ |h(t, \tau)|^2 \,
\end{equation}
that characterizes how received power is distributed across different propagation delays at given time $t$.

\subsection{RMS delay and Doppler spreads}
The \ac{RMS} delay spread is a key parameter characterizing the temporal dispersion of a wireless channel. It is given as the standard deviations of the respective power distributions, computed directly from their first and second moments. It can be obtained from \ac{PDP} for a given time $t$ and $i$-th \acp{MPC} \cite{goldsmith2005wireless}
\begin{equation}
\sigma_\tau(t) = \sqrt{ 
\frac{\sum_{i} \tau_i^2 P(t,\tau_i)}{\sum_{i} P(t,\tau_i)} 
- \left[ \frac{\sum_{i} \tau_i P(t,\tau_i)}{\sum_{i} P(t,\tau_i)} \right]^2 }.
\label{eq:rms_delay}
\end{equation}
The resulting mean value shown in Table \ref{tab:parameters} is then obtained by averaging \( \sigma_\tau(t) \) over all times.

To get the Doppler spread that characterizes the frequency dispersion of a wireless channel caused by relative motion between the transmitter and receiver, we need to first define delay-Doppler spectrum  $S(\tau, \nu)$ also known as Doppler spreading function. It can be computed by applying the discrete Fourier transform along the time dimension for each fixed delay $\tau$  
\begin{equation}
S(\tau, \nu) = \left| \mathrm{FFT}_t \left\{ h(t, \tau) \right\} \right|^2.
\label{eq:DD_Spect}
\end{equation}

Then the \ac{RMS} Doppler spread is given similarly to \eqref{eq:rms_delay} by the relation
\begin{equation}
\sigma_\nu(\tau) = \sqrt{ 
\frac{\sum_{k} \nu_k^2 S(\tau,\nu_k)}{\sum_{k} S(\tau,\nu_k)} 
- \left[ \frac{\sum_{k} \nu_k S(\tau,\nu_k)}{\sum_{k} S(\tau,\nu_k)} \right]^2 }.
\label{eq:rms_doppler}
\end{equation}
Two methods were used to estimate the \ac{RMS} Doppler spread. In the first method (Method 1) the \ac{RMS} Doppler spread is computed from the entire delay-Doppler spectrum \( S(\tau, \nu) \) according \eqref{eq:rms_doppler}. The mean RMS Doppler spread shown in Table~\ref{tab:parameters} is then obtained by averaging \( \sigma_\nu(\tau) \) over all delays. 

In the second method (Method 2) we calculate the delay Doppler spectra from \( h(\tau, t) \) within a sliding time window with a certain width and a certain window shift step. In other words, a short-term Fourier transform (STFT) is applied to the time evolution of the channel impulse response \( h(\tau, t) \). For each time window the \ac{RMS} Doppler spread \( \sigma_\nu(\tau) \) is calculated as with the Method 1.
The final mean value is the average over all time windows. This second method shows us how the \ac{RMS} Doppler spread develops during vehicle movement and is more illustrative. The disadvantage is that the window width limits the frequency resolution of the spread.

\begin{figure} [t]
    \centering
    \includegraphics[width=0.965\linewidth]{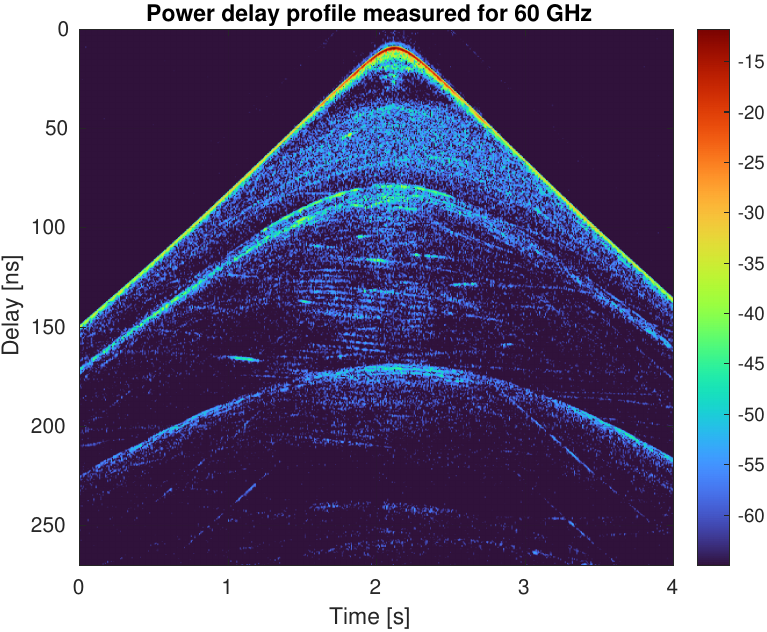}
    \caption{Power delay profile of \ac{V2V} channel measured in the 60 GHz band}
    \label{fig:PDP_60}
    \vskip -2pt
\end{figure}

\begin{figure} [t]
    \centering
    \includegraphics[width=0.965\linewidth]{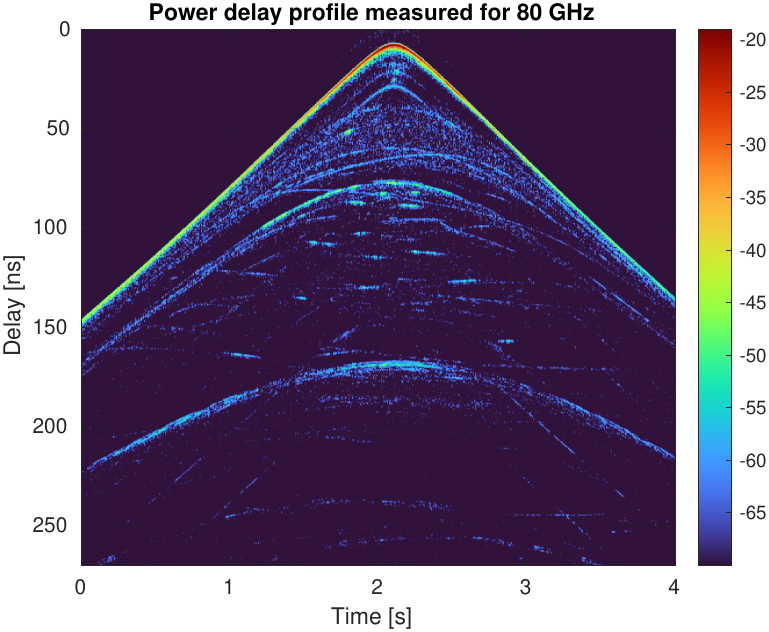}
    \caption{Power delay profile of \ac{V2V} channel measured in the 80 GHz band}
    \label{fig:PDP_80}
    \vskip-5pt
\end{figure}

\begin{figure}[b]
    \centering
    \vspace{-15pt}  
    \includegraphics[width=0.95\linewidth]{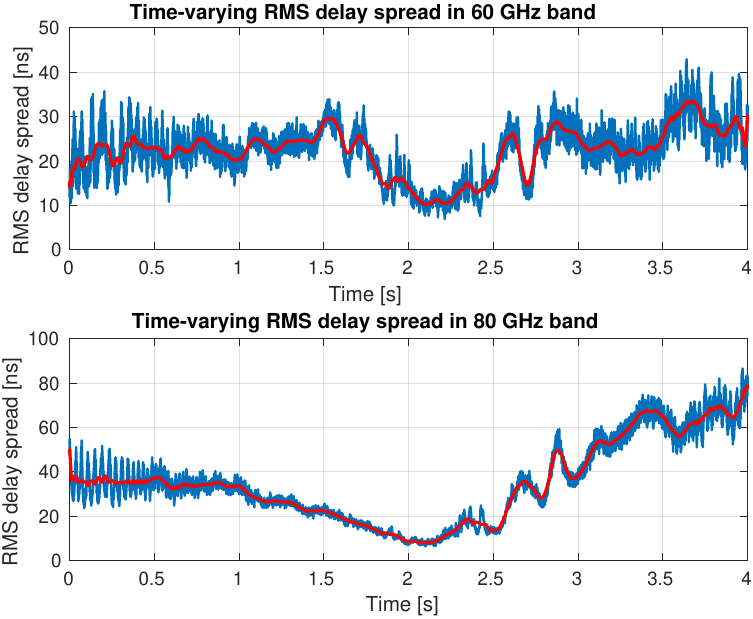}
    \caption{Time-varying RMS delay spread in the 60 GHz band (top) and 80 GHz band (bottom)}
    \label{fig:RMS_delay_60_80}
\end{figure} 



\subsection{Stationarity regions}
Stationarity regions can be identified by computing the Pearson correlation coefficient between adjacent \acp{PDP} or between \acp{PDP} at a fixed step size and tracked when the correlation drops below a predefined threshold (e.g., 0.9) \cite{Molisch_millimeter}. This approach estimates the length of quasi-stationary intervals where channel statistics remain approximately constant. The step size must be chosen to account for the high temporal resolution of the measurements, ensuring sufficient channel variation between compared \acp{PDP}. A similar approach uses a sliding window in which CIR values are averaged and correlations are made between the averages. Another option for stationarity evaluation is the \ac{LSF}, which describes spectral power variations in the time–frequency domain. Stationarity can also be assessed through collinearity in time or frequency \cite{bernado2012validity} or through the spectral distance of \acp{LSF}, using similarity thresholds. Also \ac{WSS} tests that compare second-order statistics \cite{Umansky_2009} can be applied. For comparison purposes, we used the correlation of \ac{CIR} taken at fixed intervals (every 50th \ac{CIR}).

\vskip 11pt

\begin{figure}[t]
    \centering
        \includegraphics[width=0.96\linewidth]{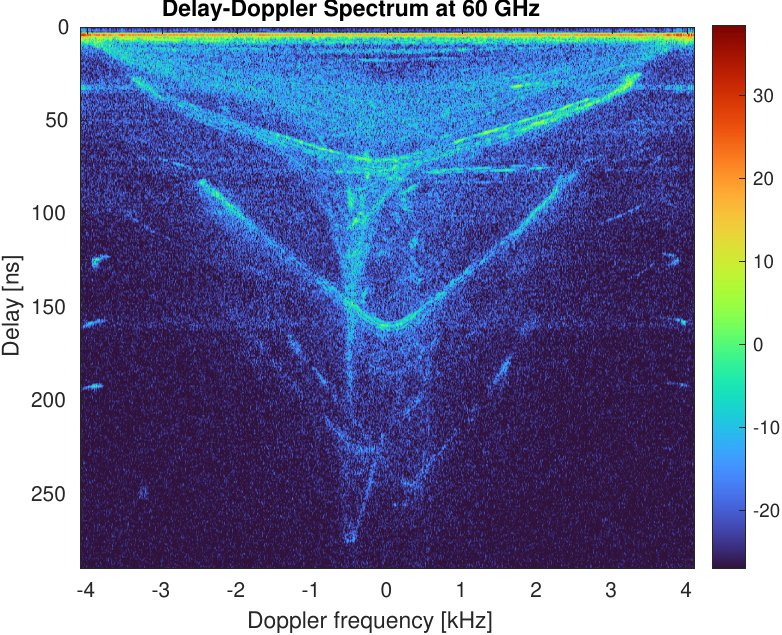}
        \caption{Delay-Doppler spectrum for 60 GHz band}
        \label{fig:DD_Spect_60}
   \vskip -3pt
\end{figure}

\section {Measurement and calculation results}
Before measurements, the channel sounder was calibrated by inserting an attenuator between the receiver and transmitter and recording the calibrated data for both frequency bands. The calibration data were then used for \ac{FMCW} processing. The measured \acp{PDP} values in both bands do not differ significantly, as can be seen in Fig.~\ref{fig:PDP_60} and Fig.~\ref{fig:PDP_80}. For the calculation, we used only a 4-second data segment from the total 8-second recording, where the multipath components have sufficient power. The delay range corresponds to a distance of approximately 0–80 m. The measured data was truncated below the noise threshold, which is about -70~dBm for both the bands. The figures show that the range between the noise peaks and the maximum value of the \ac{LOS} component is approximately 45 dB, which corresponds to the dynamic range of the receiver. It is worth mentioning that the comparative measurements of the \acp{PDP} can be affected by different antenna vertical beam-width in 60 GHz band (see Tab. \ref{tab:components}) which can create different number of \acp{MPC}. Variations in gains (listed in the horizontal plane) must also be considered.

To ensure a consistent and physically meaningful computation of the \ac{RMS} delay spread, the \acp{CIR} are time-aligned such that the \ac{LOS} component appears at a fixed delay (typically $\tau$~=~0, or close to 0) across all time instances. This compensates for the varying propagation distance due to vehicle motion and avoids artificial inflation of the delay spread caused by shifting LOS delays. This approach corresponds to a situation where the receiver is synchronized with the incoming LOS component. Calculated \ac{RMS} delay spreads are shown in Fig.~\ref{fig:RMS_delay_60_80}. Note that all one-dimensional plots in the text use blue color to display the described functional dependence and red color to show the trend obtained by a moving average filter.

\begin{figure}[t]
    \centering
        \includegraphics[width=0.95\linewidth]{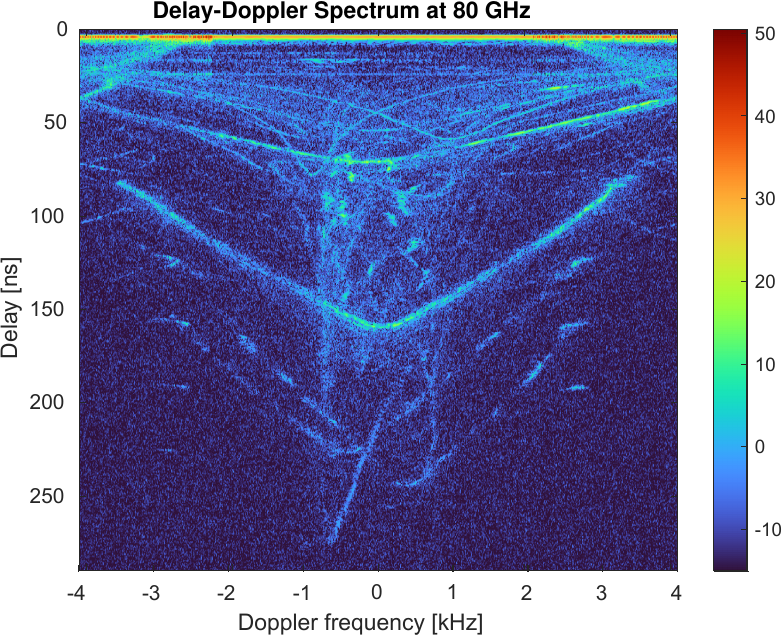}
        \caption{Delay-Doppler spectrum for 80 GHz band}
        \label{fig:DD_Spect_80}
\end{figure} 

The 80 GHz band exhibits greater dispersion. Although there appears to be a larger number of \acp{MPC} in the 60 GHz band, there is no greater dispersion of delays, probably due to their lower power. This can be observed, for example, below the LOS maximum (below the peak of the curve). It is also clear that some MPCs are present at 60 GHz when the cars are closest to each other. This is probably due to reflection from the car body, which is irradiated over a larger area due to the large beam width in the vertical direction, which is 130\degree. The RMS delay spreads for both bands can also be compared using the mean and standard deviation (std) values listed in Table~\ref{tab:parameters}.

\begin{figure}[b]
    \centering
    \vspace{-10pt}
    \includegraphics[width=0.95\linewidth]{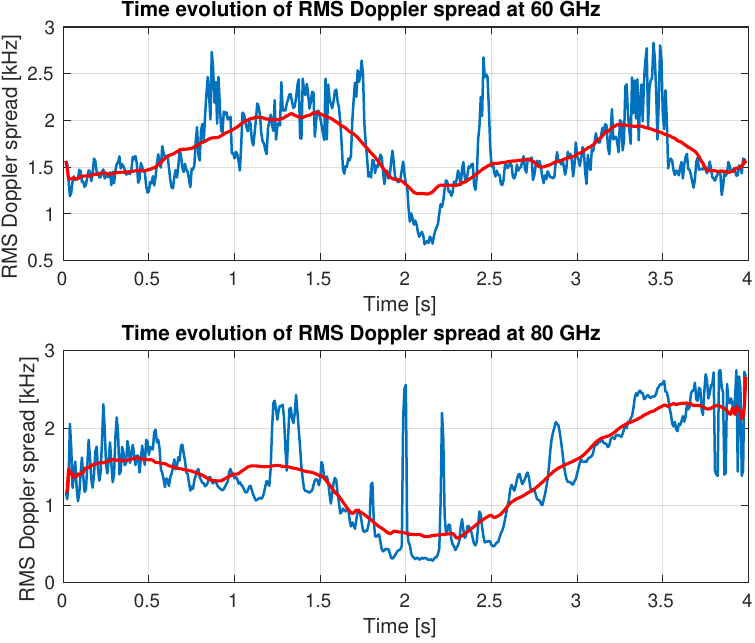}
    \caption{Time-varying RMS Doppler spread for 60 GHz band (top) and 80~GHz band (bottom) }
    \label{fig:RMS_Doppler_60_80}
\end{figure}

In the next step, we calculated the delay-Doppler spectra according to \eqref{eq:DD_Spect} using the aligned \ac{CIR}. These spectra, shown in Fig.~\ref{fig:DD_Spect_60} and Fig.~\ref{fig:DD_Spect_80}, exhibit a similar structure, with higher Doppler spread evident at 80 GHz due to a more dispersed energy in the higher Doppler bins. In general, a higher carrier frequency should also cause higher Doppler shifts. This assumption is confirmed by the calculation of the \ac{RMS} Doppler spread according to Method 1 (M1), as shown in Table \ref{tab:parameters}.

However, in order to get an idea of the changes in \ac{RMS} Doppler spread  during vehicle movement, we calculated the \ac{RMS} Doppler spread according to Method 2. We chose a window length of 256 frequency bins and a step of 64 frequency bins. The result of the calculation is shown in Fig. \ref{fig:RMS_Doppler_60_80}.
Here, the \ac{RMS} Doppler spread for the lower frequency of the transmitted signal is higher, which does not correspond to the physical nature. The discrepancy between the two methods arises probably from their different sensitivities to signal quality and antenna characteristics. The Method 1 captures the full structure of the channel in both delay and Doppler domains and is thus more robust to variations in signal amplitude and angular coverage. In contrast, the STFT-based method relies on the time evolution of individual delay taps, which makes it more sensitive to the dynamic range of the receiver and the antenna beam pattern. Weak or undetected taps due to low SNR can lead to an underestimation of the Doppler spread. 



\begin{table} [t]
\vspace{5mm} 
\centering
\caption{Selected channel parameters for two measurements}
\begin{tabular}{llcc}
\toprule
\multicolumn{4}{l}{\textbf{Measurement 1: Passing cars at the point shown in Fig.~1}} \\
\midrule
\textbf{Parameter} & & \textbf{60 GHz} & \textbf{80 GHz} \\
\midrule
\multirow{2}{*}{RMS delay spread [ns]} & mean & 18.7 & 34.9 \\
                                  & std  & 5.94 & 18.2 \\
\multirow{2}{*}{RMS Doppler spread M1 [kHz]} & mean & 1.83 & 2.05 \\
                                             & std  & 0.44 & 0.45 \\
\multirow{2}{*}{RMS Doppler spread M2 [kHz]} & mean & 1.65 & 1.44 \\
                                             & std  & 0.39 & 0.63 \\
\multirow{2}{*}{Stationarity region [s]} & mean & 0.51 & 0.48 \\
                                             & std  & 0.30 & 0.23 \\
\midrule
\multicolumn{4}{l}{\textbf{Measurement 2: Passing cars at a point 20 m farther away}} \\
\midrule
\textbf{Parameter} & & \textbf{60 GHz} & \textbf{80 GHz} \\
\midrule
\multirow{2}{*}{RMS delay spread [ns]} & mean & 22.1 & 38.9 \\
                                  & std  & 5.43 & 14.7 \\
\multirow{2}{*}{RMS Doppler spread M1 [kHz]} & mean & 1.91 & 2.16 \\
                                             & std  & 0.48 & 0.45 \\
\multirow{2}{*}{RMS Doppler spread M2 [kHz]} & mean & 1.81 & 1.57 \\
                                             & std  & 0.45 & 0.55 \\
\multirow{2}{*}{Stationarity region [s]} & mean & 0.47 & 0.42 \\
                                             & std  & 0.27 & 0.21 \\
\bottomrule
\end{tabular}
\label{tab:parameters}
\vspace{-5mm}
\end{table}

It should also be noted that the calculation result is influenced by both the window width and the step size. A step size of about 25 \% of the window length offers a good compromise between temporal resolution and estimation variance for the considered \ac{V2V} scenarios. It mainly affects the smoothness of the results: smaller values mean larger overlaps and more frequent RMS Doppler spread sampling. While mean values stay almost unchanged, the estimates are more stable, and the time evolution smoother. With shorter windows, RMS Doppler spread tends to be overestimated due to limited frequency resolution, whereas longer windows converge to the global estimate of Method 1. Note that changing the window length from 64 bins to 1024 bins causes the RMS Doppler spread mean value to change by only about 13\%.

Due to the high speed of CIR measurement and the resulting high correlation of adjacent snapshots, a period of 50 CIR was used to evaluate stationarity regions, which set the measurement interval resolution to 6.25 ms. The Table \ref{tab:parameters} clearly shows that the dispersion of values for stationarity regions is relatively large, which is to be expected given the nature of the masured \acp{PDP}. The method used is quite sensitive to the choice of parameters, but for the purpose of comparing the two bands it can serve well.

\section{Conclusions}

A comparison of the 60 GHz and 80 GHz bands leads to the conclusion that the two bands do not differ significantly. This is evident from all graphs. Although short-term changes are often different, their trends are similar. It is clear that both RMS delay and Doppler spreads are smallest when the two cars pass each other. The delay spread at this moment is mainly defined by a strong LOS component, and the Doppler shifts are minimal because the relative velocity is zero at the moment of encounter. 
It is generally known that a higher \ac{RMS} Doppler spread leads to shorter time intervals over which the channel remains approximately stationary. This is confirmed by the results shown in the table (\ac{RMS} Doppler spread vs. stationarity regions). Rapidly changing \ac{RMS} delay spread leads to short stationarity regions. This is difficult to assess, as the rates of change in Fig.~\ref{fig:RMS_delay_60_80} appear to be identical. Certain differences in the results can also be attributed to the unequal radiation characteristics of the antennas in the measurement setup. 
The different \ac{RMS} Doppler spread values obtained by Method 1 and Method 2 have been explained and are apparently caused by a shorter data segment for calculation and thus greater sensitivity to noise and other system parameters.
The results of the Measurement 1 and 2 shown in Table \ref{tab:parameters} show some minor differences resulting from larger number of reflective objects in the second measurement leading to a higher RMS delay spread. It's also likely that the relative speed of the vehicles was in reality slightly higher. 

\section*{Acknowledgement}
The research described in this paper was financed by the Czech Science Foundation, Project No. 23-04304L, ”Multiband prediction of millimeter-wave propagation effects for dynamic and fixed scenarios in rugged time varying environments” and by the National Science Centre, Poland, Project No. 2021/43/I/ST7/03294, through the OPUS-22 (LAP) Call in the Weave Program.

\balance
\bibliographystyle{IEEEtran} 
\bibliography{papers} 

@book{goldsmith2005wireless,
  author    = {Andrea Goldsmith},
  title     = {Wireless Communications},
  year      = {2005},
  publisher = {Cambridge University Press},
  address   = {New York, USA},
  isbn      = {9780521837163}
}

@INPROCEEDINGS{Dupleich_Multi-Band_T,
  author={Dupleich, Diego and Muller, Robert and Schneider, Christian and Skoblikov, Sergii and Luo, Jian and Boban, Mate and Del Galdo, Giovanni and Thoma, Reiner},
  booktitle={2019 IEEE 2nd Connected and Automated Vehicles Symposium (CAVS)}, 
  title={Multi-Band Vehicle to Vehicle Channel Measurements from 6 {GHz} to 60 {GHz} at "{T}" Intersection}, 
  year={2019},
  volume={},
  number={},
  pages={1-5},
  doi={10.1109/CAVS.2019.8887794}}

@INPROCEEDINGS{Dupleich_Multi-Band,
  author={Dupleich, Diego and Müller, Robert and Skoblikov, Sergii and Schneider, Christian and Luo, Jian and Boban, Mate and Del Galdo, Giovanni and Thomä, Reiner},
  booktitle={2018 IEEE 29th Annual International Symposium on Personal, Indoor and Mobile Radio Communications (PIMRC)}, 
  title={Multi-band Characterization of Path-loss, Delay, and Angular Spread in {V2V} Links}, 
  year={2018},
  volume={},
  number={},
  pages={85-90},
  doi={10.1109/PIMRC.2018.8580797}}

@INPROCEEDINGS{Takahashi_Distance,
  author={Takahashi, S. and Kato, A. and Sato, K. and Fujise, M.},
  booktitle={2003 IEEE 58th Vehicular Technology Conference. VTC 2003-Fall (IEEE Cat. No.03CH37484)}, 
  title={Distance dependence of path loss for millimeter wave inter-vehicle communications}, 
  year={2003},
  volume={1},
  number={},
  pages={26-30 Vol.1},
  doi={10.1109/VETECF.2003.1284971}}

@INPROCEEDINGS{Hoellinger_V2V,
  author={Hoellinger, Joseph and Makhoul, Gloria and D’Errico, Raffaele and Marsault, Thierry},
  booktitle={2022 International Symposium on Antennas and Propagation (ISAP)}, 
  title={{V2V} Dynamic Channel Characterization in {5G} mm{W}ave Band}, 
  year={2022},
  volume={},
  number={},
  pages={525-526},
  doi={10.1109/ISAP53582.2022.9998590}}

@INPROCEEDINGS{Chopra_A_Real_Time,
  author={Chopra, Aditya and Thornburg, Andrew and Kanhere, Ojas and Ghassemzadeh, Saeed S. and Majmundar, Milap and Rappaport, Theodore S.},
  booktitle={2022 IEEE Wireless Communications and Networking Conference (WCNC)}, 
  title={A Real-Time Millimeter Wave {V2V} Channel Sounder}, 
  year={2022},
  volume={},
  number={},
  pages={2607-2612},
  doi={10.1109/WCNC51071.2022.9772001}}

@ARTICLE{Ghosh_V2V,
  author={Ghosh, Anirban and Chandra, Aniruddha and Mikulasek, Tomas and Prokes, Ales and Wojtun, Jaroslaw and Kelner, Jan M. and Ziolkowski, Cezary},
  journal={IEEE Antennas and Wireless Propagation Letters}, 
  title={Vehicle-to-Vehicle Path Loss Modeling at Millimeter-Wave Band for Crossing Cars}, 
  year={2023},
  volume={22},
  number={9},
  pages={2125-2129},
  doi={10.1109/LAWP.2023.3277961}}

@ARTICLE{Liu_Measurements,
author={Liu, Xichen and Liu, Shuning and Zhou, Wangyang and Yang, Lin and Yue, Guangrong},
journal={IEEE Transactions on Wireless Communications}, 
title={Measurements and Modeling of Millimeter-Wave Vehicle-to-Vehicle Propagation With Vehicle Obstructions}, 
year={2025},
volume={24},
number={5},
pages={4024-4039},
doi={10.1109/TWC.2025.3540968}}

@InProceedings{Wang_fading,
author="Wang, Hui
and Yin, Xuefeng
and Cai, Xuesong
and Wang, Haowen
and Yu, Ziming
and Lee, Juyul",
title="Fading Characterization of 73 {GHz} Millimeter-Wave {V2V} Channel Based on Real Measurements",
booktitle="Communication Technologies for Vehicles",
year="2018",
publisher="Springer International Publishing",
address="Cham",
pages="159--168",
}

@INPROCEEDINGS{Umansky_2009,
  author={Umansky Dmitry and Patzold, Matthias},
  booktitle={GLOBECOM 2009 - 2009 IEEE Global Telecommunications Conference}, 
  title={Stationarity Test for Wireless Communication Channels}, 
  year={2009},
  volume={},
  number={},
  pages={1-6},
  doi={10.1109/GLOCOM.2009.5425841}
}

@ARTICLE{Molisch_millimeter,
  author={Molisch, Andreas F. and Mecklenbräuker, Christoph F. and Zemen, Thomas and Prokes, Ales and Hofer, Markus and Pasic, Faruk and Hammoud, Hussein},
  journal={IEEE Open Journal of Vehicular Technology}, 
  title={Millimeter-Wave {V2X} Channel Measurements in Urban Environments}, 
  year={2025},
  volume={6},
  number={},
  pages={520-541},
  doi={10.1109/OJVT.2024.3521637}}

@ARTICLE{Zavorka_Charact,
  author={Zavorka, Radek and Mikulasek, Tomas and Vychodil, Josef and Blumenstein, Jiri and Chandra, Aniruddha and Hammoud, Hussein and Kelner, Jan M. and Ziółkowski, Cezary and Zemen, Thomas and Mecklenbräuker, Christoph and Prokes, Ales},
  journal={IEEE Access}, 
  title={Characterizing the 80 {GHz} Channel in Static Scenarios: Diffuse Reflection, Scattering, and Transmission Through Trees Under Varying Weather Conditions}, 
  year={2024},
  volume={12},
  number={},
  pages={144738-144749},
  doi={10.1109/ACCESS.2024.3472003}}

@article{hammoud2024TVT,
	author = {H. Hammoud and Y. Zhang and Z. Cheng and S. Sangodoyin and M. Hofer and F. Pasic and T. Pohl and R. Z{\'a}vorka and A. Proke{\v s} and T. Zemen and C. F. Mecklenbr{\"a}uker and A. F. Molisch},
	journal = {arxiv:2412.01165},
	publisher = {IEEE},
	title = {Double-Directional {V2V} Channel Measurement
using ReRoMA at 60 {GHz}},
	year = {2024}}

@article{boban2019multi,
	author = {Boban, Mate and Dupleich, Diego and Iqbal, Naveed and Luo, Jian and Schneider, Christian and M{\"u}ller, Robert and Yu, Ziming and Steer, David and J{\"a}ms{\"a}, Tommi and Li, Jian and others},
	journal = {IEEE Access},
	pages = {9724--9735},
	publisher = {IEEE},
	title = {Multi-band vehicle-to-vehicle channel characterization in the presence of vehicle blockage},
	volume = {7},
	year = {2019}}

@inproceedings{groll2019sparsity,
	address = {Shanghai, China},
	author = {Groll, Herbert and others},
	booktitle = {2019 IEEE International Conference on Communications Workshops (ICC Workshops)},
	doi = {10.1109/ICCW.2019.8756930},
	month = {May},
	organization = {IEEE},
	title = {Sparsity in the delay-{Doppler} domain for measured 60 {GHz} vehicle-to-infrastructure communication channels},
	year = {2019}}

@article{yamamoto2008path,
	author = {Yamamoto, Atsushi and Ogawa, Koichi and Horimatsu, Tetsuo and Kato, Akihito and Fujise, Masayuki},
	journal = {IEEE Transactions on Vehicular Technology},
	number = {1},
	pages = {65--78},
	publisher = {IEEE},
	title = {Path-loss prediction models for intervehicle communication at 60 {GHz}},
	volume = {57},
	year = {2008}}

@inproceedings{bernado2012validity,
	author = {Bernad{\'o}, Laura and Zemen, Thomas and Tufvesson, Fredrik and Molisch, Andreas F and Mecklenbr{\"a}uker, Christoph F},
	booktitle = {IEEE 23rd International Symposium on Personal, Indoor and Mobile Radio Communications (PIMRC)},
	organization = {IEEE},
	pages = {1757--1762},
	title = {The (in-) validity of the WSSUS assumption in vehicular radio channels},
	year = {2012}}

@article{vychodil2019iet,
	author = { Vychodil, Josef and Pospisil, Martin and Prokes, Ales and Blumenstein, Jiri },
	journal = {IET Communications},
	number = {3},
	pages = {331--338},
	publisher = {IET},
	title = {Millimetre wave band time domain channel sounder},
	volume = {13},
	year = {2019}}

\end{document}